\documentclass[12pt,preprint]{aastex}  
\usepackage{natbib}
\usepackage[usenames,dvips]{color}

\shorttitle{Analytical expressions for the envelope binding energy}
\shortauthors{Loveridge, van der Sluys \& Kalogera}

\newcommand{\msun}{M_\odot}
\newcommand{\rsun}{R_\odot}

\newcommand{\Ebind}{E_\mathrm{bind}}
\newcommand{\lam}{\lambda_\mathrm{env}}
\newcommand{\eg}{\emph{e.g.}}
\newcommand{\ie}{\emph{i.e.}}

\begin{document}

\title{Analytical expressions for the envelope binding energy of giants\\
  as a function of basic stellar parameters}


\author{A.J.\ Loveridge}
\affil{Department of Physics and Astronomy, Northwestern University, 2131 Tech Drive, Evanston, IL 60208}

\author{M.V.\ van der Sluys\altaffilmark{1,2,3}}
\affil{Department of Physics, University of Alberta, 11322 - 89 Ave, Edmonton AB, T6G 2G7, Canada}
\altaffiltext{1}{CITA National Fellow}
\altaffiltext{2}{Department of Physics and Astronomy, Northwestern University, 2131 Tech Drive, Evanston, IL 60208}
\altaffiltext{3}{Now at Department of Astrophysics, Radboud University Nijmegen, PO Box 9010, NL-6500 GL Nijmegen, the Netherlands}

\and

\author{V.\ Kalogera}
\affil{Department of Physics and Astronomy, Northwestern University, 2131 Tech Drive, Evanston, IL 60208}

\begin{abstract}
  The common-envelope (CE) phase is an important stage in the evolution of binary stellar populations.  The most common
  way to compute the change in orbital period during a CE is to relate the binding energy of the envelope
  of the Roche-lobe filling giant to the change in orbital energy.  Especially in population-synthesis 
  codes, where the evolution of millions of stars must be computed and detailed evolutionary models are too
  expensive computationally, simple approximations are made for the envelope binding energy.
  In this study, we present accurate analytic prescriptions based on detailed stellar-evolution models 
  that provide the envelope binding energy for giants with metallicities between $Z=10^{-4}$ and $Z=0.03$
  and masses between $0.8\,\msun$ and $100\,\msun$, as a function of the metallicity, mass, radius and 
  evolutionary phase of the star.  Our results are also presented in the form of electronic data tables
  and \texttt{Fortran} routines that use them.  We find that the 
  accuracy of our fits is better than 15\% for 90\% of our model data points in all cases, and better than 
  10\% for 90\% of our data points in all cases except the asymptotic giant branches for three of the six metallicities we consider.
  For very massive stars ($M\gtrsim50\,M_\odot$), when 
  stars lose more than $\sim20\%$ of their initial mass due to stellar winds, our fits do not describe the 
  models as accurately.  Our results are more widely applicable --- covering wider ranges of metallicity and mass ---
  and are of higher accuracy than those of previous studies.
\end{abstract}

\keywords{stars: evolution, stars: fundamental parameters, stars: mass loss, binaries: close}

\section{Introduction}
\label{sec:intro}

The common-envelope (CE) phase \citep{1976IAUS...73...75P,1984ApJ...277..355W,2000ARA&A..38..113T} is an important event in the 
evolution of many binaries and is used to explain
large numbers of observed compact binaries, such as X-ray binaries, cataclysmic variables (CVs) and double degenerates,
as well as bipolar planetary nebulae \citep{1990ApJ...358..189D}.
The CE phase is believed to be initiated when an evolved, giant star is in orbit with a more compact companion
and fills its Roche lobe.  When such a donor star has a deep convective envelope and/or the mass ratio of the
system is sufficiently large, unstable mass transfer occurs \citep{1996ApJ...471..366R}, resulting in a fast expanding envelope
which quickly engulfs the companion.  Inside this common envelope, the binary orbit of the core of the giant and the secondary star
shrinks due to friction and torques, leading to either a compact binary consisting of 
these two objects or their merger.
The energy generated through orbital shrinkage heats and eventually expels
the envelope.  The whole process is thought to occur on time scales much shorter than the nuclear-evolution 
timescales of stars ($\lesssim\,10^3$\,yr) so that the mass of the giant's core does not change, and the secondary
is not affected \citep{2000ARA&A..38..113T}.

Despite the fact that CEs are an important ingredient for modelling stellar systems and populations, 
the details of the process are poorly understood.  Three-dimensional hydrodynamical models have provided detailed
simulations for the first month or so of CE evolution, in which significant orbital shrinkage takes place, but 
due to the large differences in scale between a red-giant envelope and its core, the precise outcome of the CE 
phase cannot yet be predicted \citep[\eg,][]{2006astro.ph.11043T}.  Instead, a very simplified scenario 
is commonly used, where the binding energy is equated to the difference in orbital energy before and after the CE in order 
to predict the post-CE orbital period \citep{1984ApJ...277..355W}.  Part of the uncertainty in CE evolution is 
allowed for in the CE parameter $\alpha_\mathrm{CE}$, which specifies the efficiency with which
orbital energy is used to expel the envelope.  Especially for population-synthesis codes, 
where the evolution of millions of stars must be computed and models are very basic, the stellar structure needed to 
compute the envelope binding energy is not available and the envelope-structure parameter $\lam$ 
(see Eq.\,\ref{eq:lambda} and the discussion in Sect.\,\ref{sec:lambda})
is used to approximate the binding energy from basic stellar parameters.  In many studies, this parameter has 
been assumed constant, typically $\lam \approx 0.5$ \citep[\eg,][]{1987A&A...183...47D,
2000A&A...360.1011N,2002MNRAS.329..897H}, or $\alpha_\mathrm{CE} \lam = 1.0$ or $0.5$ 
\citep{2008ApJS..174..223B}, whereas in reality it can vary wildly during the evolution of a star, especially on 
the asymptotic giant branch (AGB) \citep[\eg,][]{2000A&A...360.1043D,2006A&A...460..209V}.  In fact, 
\citet{MvdS_Mykonos} show that the latter assumption implies $\alpha_\mathrm{CE} > 1$ for about 60\% of the
CEs that occur in a stellar population of solar metallicity.
Note that an alternative scenario for envelope ejection, based on angular-momentum balance 
\citep{2000A&A...360.1011N,2005MNRAS.356..753N,2006A&A...460..209V}, does not
depend on the envelope binding energy and is therefore not affected by our results.

In this paper, we use the stellar-evolution code \texttt{ev} \citep[\eg,][and references therein]{2005ApJ...629.1055Y} 
to compute detailed stellar-evolution models for a range of masses ($0.8-100\,M_\odot$) and six different
metallicities, and compute their envelope 
binding energies throughout their evolution.  We fit the binding energies as a function of basic stellar 
parameters (metallicity, mass and radius) and provide the results of these fits, as well as routines to compute the envelope
binding energy for any combination of mass and radius, for a given metallicity in our grid.  We find that our analytic prescriptions
describe 90\% of our model data points with an accuracy better than 10--15\% for the metallicities provided.
Recently, \citet{2010ApJ...716..114X} have used the same approach, providing fits for 14 discrete masses
and two different metallicities.  Our study improves on that by using the stellar mass as a fitting parameter,
so that any mass --- and masses from a wider range --- can be used, and by providing a larger number and wider range of metallicities.  
In addition, we show that our fits are largely independent of the wind mass loss, allowing for many wind prescriptions, 
and we provide a more detailed analysis of our accuracies.

In Section~\ref{sec:evolution} of this paper, we present the
stellar-evolution code used and the models generated with it.  Section~\ref{sec:fitting} describes the method we
use to fit the binding energies and in Section~\ref{sec:results} we present our results and their accuracies.
Section~\ref{sec:discussion} contains a discussion of this study and we present a summary and conclusions in
Section~\ref{sec:conclusions}.

\section{Stellar evolution}
\label{sec:evolution}

Here we describe the stellar-evolution code that we use for this study, the relevant details of the physics
involved, as well as the actual grids of detailed stellar-evolution models that are computed to perform our
fits and determine their accuracy.

\subsection{Stellar-evolution code}
\label{sec:code}

We compute our stellar models using a version of the binary stellar evolution code \texttt{ev} (also known as 
\texttt{STARS} or \texttt{TWIN})\footnote{The current version of \texttt{ev} is obtainable on request from 
\texttt{eggleton1@llnl.gov}, along with data files and a user manual.}, developed by Eggleton 
\citep[][and references therein]{1971MNRAS.151..351E, 1972MNRAS.156..361E,2005ApJ...629.1055Y} and updated as 
described in \citet{1995MNRAS.274..964P}.  In the code, convective mixing is modelled by a diffusion equation with a
ratio of mixing length to pressure scale height of $l/H_\mathrm{p} = 2.0$.  Convective overshooting in the core is taken 
into account on the main sequence (MS) for stars with $M > 1.2\,M_\odot$ and on the horizontal branch (HB) for stars 
of all masses.  On the MS, we use an overshooting parameter $\delta_\mathrm{ov}=0.12$ for stars with $M > 2.0\,M_\odot$, 
which corresponds to an overshooting length of about $0.3\,H_\mathrm{p}$, and $0.0 < \delta_\mathrm{ov} < 0.12$ 
increasing linearly with mass for stars with $1.2\,M_\odot < M < 2.0\,M_\odot$.  On the HB, we use $\delta_\mathrm{ov}=0.12$
for all models.  The code cannot evolve a model through the helium flash, the violent ignition of helium in a low-mass
star with a degenerate core, but it automatically replaces the model at the moment of helium ignition with a tailored 
model of the same total mass and core mass in which helium has just ignited.  For stars above a 
certain mass ($M \gtrsim 2.1\,M_\odot$ for $Z=0.02$), helium ignites non-degenerately and this intervention is not needed.

We define the helium-core mass $M_\mathrm{c}$ as the mass coordinate below which the hydrogen abundance does 
not exceed 10\%.  We compute the binding energy of the hydrogen-rich envelope, $\Ebind$, by integrating the 
gravitational and internal energies over the mass coordinate of the model, from the core-envelope boundary to the 
surface of the star $M_\mathrm{s}$:
\begin{equation}
  \Ebind = \int_{M_\mathrm{c}}^{M_\mathrm{s}} E_\mathrm{int}(m) - \frac{G m}{r(m)}\, \mathrm{d}m
  \label{eq:Ebind}
\end{equation}
The term $E_\mathrm{int}$ is the internal energy per unit of mass, which contains 
terms such as the thermal energy of the gas and the radiation energy, but not the recombination energy.
More details regarding these assumptions are provided in \cite{2006A&A...460..209V}.  

In this study, we compute grids of models with six different metallicities: 
$Z = 10^{-4}$, $Z = 0.001$, $Z = 0.010$, $Z = 0.015$, $Z = 0.020$, and $Z = 0.030$.
The initial hydrogen and helium abundances ($X$ and $Y$, respectively) 
of our model stars are a function of the metallicity $Z$, and are given by $X = 0.76 - 3.0Z$ and $Y = 0.24 + 2.0Z$.  

For three of these metallicities, $Z = 10^{-4}$, $Z = 0.020$, and $Z = 0.030$, we calculate additional grids that 
include mass loss via stellar winds.  We use a prescription that was inspired by \citet{1975MSRSL...8..369R}:
\begin{equation}
  \dot{M} = - \eta \cdot \min\left\{ 
    \begin{array}{l}
      3.16 \cdot 10^{-14}\,M_\odot\,\mathrm{yr}^{-1}\, \left(\!\frac{M}{M_\odot}\!\right) \left(\!\frac{L}{L_\odot}\!\right) \left(\!\frac{\Ebind}{10^{50}\,\mathrm{erg}}\!\right)^{-1}  \\
      9.61 \cdot 10^{-10}\,M_\odot\,\mathrm{yr}^{-1}\, \left(\!\frac{L}{L_\odot}\!\right) 
    \end{array} 
  \right.,
  \label{eq:wind} 
\end{equation}
where we set $\eta = 0.2$ for this study.  This wind prescription dominates the stellar winds for lower-mass stars 
on the giant branches, and the upper prescription in Eq.\,\ref{eq:wind} is used in most of those models, 
except where $|\Ebind|$ is small, \eg, for stars near the tip of the asymptotic giant 
branch (AGB).  For massive, luminous stars, we use the wind prescription by \cite{1988A&AS...72..259D}.

\subsection{Stellar-evolution models}
\label{sec:models}

We computed several grids of single-star models for a range of metallicities and for each metallicity, 
for a range of initial masses.  For these grids, we assume that there is no mass loss due to stellar
winds.  However, for three metallicities ($Z=0.02$ (``solar'') and the two extremes $Z=10^{-4}$ and $Z=0.03$), 
we computed a grid of models where wind mass loss is included, in order to determine how the accuracy of our 
fits is affected by stellar winds and how we can correct for that (see Sect.\,\ref{sec:masslossprocedure} 
and \ref{sec:winds}).

We selected 73 zero-age main-sequence (ZAMS) masses between $0.8\,M_\odot$ and $100\,M_\odot$ in each 
grid, distributed uniformly in $\log M$:
\begin{equation}
  \log M_i \simeq -0.038660 + 0.029124 \cdot i
  \label{eq:masses}
\end{equation}
resulting in $M \approx 0.80, 0.86, 0.91, \ldots, 93.5, 100.0\,M_\odot$.  We ignore stars that have
a main-sequence lifetime longer than 13\,Gyr for a given metallicity.  We use the models with an
\emph{odd} value of $i$ to perform the fits, whereas the models with an \emph{even} value for $i$
are used as verification, to determine the accuracies of our fits for models that they are 
independent of.

\begin{figure}  
  \includegraphics[angle=-90,width=\columnwidth]{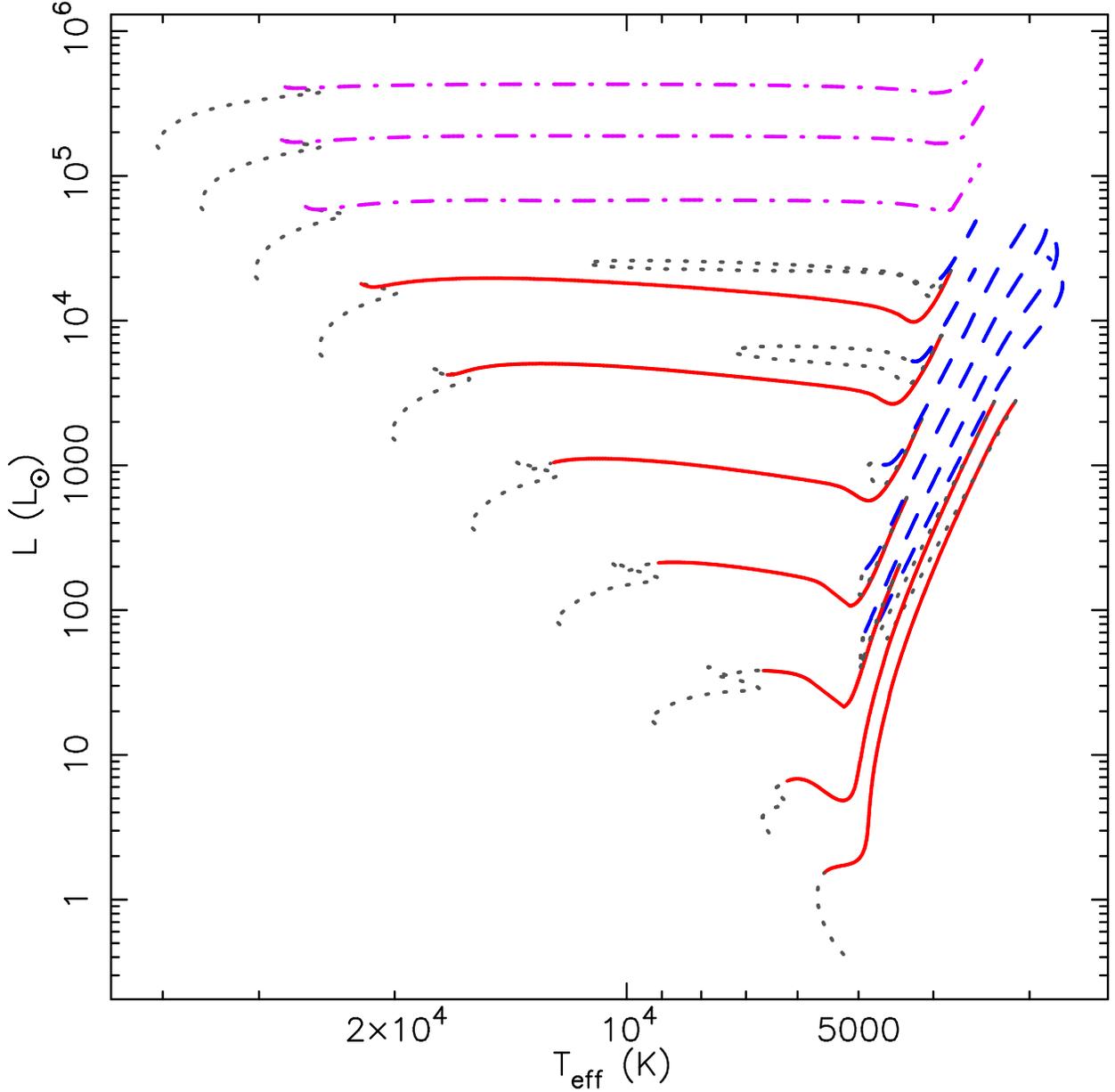}
  \figcaption{
    Hertzsprung-Russel diagram for selected models
    with masses of 0.91, 1.36, 2.04 3.05, 4.57, 6.84, 10.22, 15.3, 22.9 and 34.2\,$\msun$, 
    from our $Z=0.02$ grid.  The different line styles and colours indicate different evolutionary
    stages. 
    Dotted (grey) lines are used for the main sequence, (part of the) Hertzsprung gap and horizontal branch, phases
    where the star either has no deep convective envelope or cannot fill its Roche lobe.  The solid
    (red) lines indicate the red-giant branch (LMR), the dashed (blue) lines show the asymptotic giant 
    branch (LMA) and the dash-dotted (magenta) lines indicate the high-mass models (HM), which have only 
    one giant branch.
    \label{fig:hrd_tracks}
  }
\end{figure}

The full grids of detailed stellar-evolution models are computed for all metallicities considered.  
For the three stellar-wind grids, only odd-$i$ models are calculated, since this is all that is required
to determine the mass-loss correction factor described in Sect.\,\ref{sec:masslossprocedure}.
A Hertzsprung-Russel diagram (HRD) for selected stellar models
from our grid with $Z=0.02$ is shown in Fig.\,\ref{fig:hrd_tracks}.

\section{Method of fitting}
\label{sec:fitting}

With the data generated as explained above we construct fit functions to 
model the behavior of the binding energy. In this section, we describe the 
procedure by which these fits are obtained.  Section~\ref{sec:fitpar}
describes our choice of fitting parameters, in Sect.\,\ref{sec:divtracks}
we describe how we divide our evolutionary tracks into groups which will
be fitted separately, and Sect.\,\ref{sec:fitfuncs} contains the formulation 
of the fitting functions and a description of the fitting procedure itself.
Finally, in Sect.\,\ref{sec:masslossprocedure} we present a correction factor 
that takes into account the most important effects of wind mass loss.

\subsection{Choice of fitting parameters}
\label{sec:fitpar}

After some experimentation we find that the binding energy is best described 
by the stellar mass and radius, as opposed to other basic parameters like age or core mass.
Such a relationship of the binding 
energy with the mass and radius of a star was also found by 
\citet{2000A&A...360.1043D}.  
Because the ranges of $\Ebind$, $R$ and $M$ may span several orders 
of magnitude, we use the (base-10) logarithm of these variables.

We do our fits separately for each metallicity and do not provide fits that
take into account metallicity as a parameter.  The reason for this choice is
that with each added fitting parameter, subtle features in the models that were 
well-described by the other parameters are washed out when the new 
dimension is added.  However, we compare the changes due to the metallicity
in the $\log\Ebind-\log R$ plane for models with a given mass, and find that
these changes are relatively smooth, see Fig.\,\ref{fig:be_tracks_Zs}.
This suggests that the user could 
compute the envelope binding energy for the two metallicities that bracket
the desired value of $Z$ and interpolate linearly in $\log\Ebind$ in order
to estimate the binding energy for the given star. In practice, however, 
this grid may be sufficiently dense that the user can take the value of $Z$ 
that is closest to the desired value.

\subsection{Dividing the evolutionary tracks}
\label{sec:divtracks}

A CE phase can only be initiated by a star that is at its maximum
radius so far in its evolution.  Because of this 
we consider only evolutionary phases in which a star expands (almost) monotonously.  
This results in the division of the evolutionary tracks into four different groups, 
and we develop fits for each group separately.

First, we divide our set of stellar-evolution models into low-mass and a high-mass groups.  
The dividing mass between the two groups lies around $12\,\msun$, depends lightly on metallicity, and is 
given in Table~\ref{tab:Mlh}.

\begin{table}
  \centering
  \caption{
    Dividing mass between the low-mass (LMR/LMA) and high-mass (HM) models for each metallicity considered
    \label{tab:Mlh}
  }
  \begin{tabular}{ll}
    \multicolumn{2}{c}{$M_\mathrm{lh}$} \\
    \hline
    $Z=10^{-4}$ &  $11.7\,M_\odot$ \\
    $Z=0.001$   &  $11.7\,M_\odot$ \\
    $Z=0.01$    &  $10.2\,M_\odot$ \\
    $Z=0.015$   &  $11.7\,M_\odot$ \\
    $Z=0.02$    &  $11.7\,M_\odot$ \\
    $Z=0.03$    &  $13.4\,M_\odot$ \\
  \end{tabular}
\end{table}

The low-mass stars in our grid ($M\lesssim 12\,M_\odot$) evolve through a clear 
horizontal-branch (HB) phase between the red-giant branch (RGB) and asymptotic giant 
branch (AGB).  In this phase, the star burns helium in its core and the star's radius 
is smaller than it was on the RGB, so that this phase can be ignored for common-envelope evolution.  In addition,
most of these low-mass stars undergo a \emph{first dredge-up} on the RGB,
during which the hydrogen-burning shell meets less-processed material and 
the star as a whole shrinks slightly.  
We define the beginning of the RGB as the moment where the hydrogen abundance
in the stellar core becomes zero.  As a consequence, the envelope binding energy 
in the Hertzsprung gap (HG) is also described by the RGB fits for most stars
(see Fig.\,\ref{fig:hrd_tracks}). 
We define the end of the RGB as the point of maximum radius before the core helium abundance 
drops below half of its absolute maximum value over the stellar lifetime. The beginning of the 
AGB is taken to be the point where the core helium abundance drops below approximately 10\%
(we vary this value slightly for different stars). We find that this roughly defines
the point where the stellar radius begins to increase again after the HB.  For all of our
models, the ``AGB'' begins well before the radius surpasses its maximum on the RGB, so that we 
cover more than the range of radii for which a CE can be initiated.
As the tip of the AGB, we take the model where the star reaches its maximum
radius in its entire evolution.
We find that the accuracies of our fits are greatly improved
if we treat the tracks before and after this dredge-up phase separately.
Since the dredge-up occurs at different radii for models with different
masses, we include a simple function to describe this radius as a
function of the stellar mass:
\begin{equation}
  R_\mathrm{LMR} = \sum_{i=0}^4 a_i \left[\log \left(\frac{M}{\msun}\right)\right]^i,
  \label{eq:rgbdiv}
\end{equation}
with the coefficients $a_i$ given in Table~\ref{tab:rgbdivcoeff}.

\begin{table}
  \caption{
    Coefficients for the fit of the division between the groups LMR1 and LMR2 in Eq.\,\ref{eq:rgbdiv}.
    \label{tab:rgbdivcoeff}
  }
  \begin{tabular}{llllll}    
    $Z$ & $a_0$ & $a_1$ & $a_2$ & $a_3$ & $a_4$ \\
    \hline
    $10^{-4}$ & 0.604723 & 1.39724 & -0.0824963  & 1.14143 & 0 \\
    $0.001$ & 0.4566 & 1.1866  & 2.39088 & -3.05404  & 1.4049 \\
    $0.01$  & 0.282174 & 1.14938  & 1.88445 & -1.0823 & 0 \\
    $0.015$ & 0.251818 & 1.21049  & 1.6349 & -0.83691 & 0 \\
    $0.02$  & 0.240637 & 1.08922  & 1.95318 & -1.03218 & 0 \\
    $0.03$  & 0.234888 & 0.897294  & 2.51995 & -1.41411 & 0 \\
  \end{tabular}
\end{table}

As a result, we split the tracks from the low-mass 
models into three groups, \textbf{LMR1} for the low-mass early RGB (before the dredge-up), 
\textbf{LMR2} for the low-mass late RGB and \textbf{LMA} for the low-mass AGB.

For the models with higher masses ($M \gtrsim 12\,M_\odot$) there is no HB
and no first dredge-up, and we keep these tracks intact and put
them in a fourth group, which we call \textbf{HM}.  Hence, these tracks are
defined between the moment hydrogen is exhausted in the core near the TAMS and
the maximum radius ever achieved in the evolution of each star.  The lowest-mass model 
(whose mass is $\sim 12 \msun$, see Table~\ref{tab:Mlh})
included in this high-mass group is also included in the low-mass RGB and 
AGB groups to ensure continuity between the regions (we use the 
phenomenological parallels in the $\log \Ebind-\log R$ curves (see Fig.\,\ref{fig:be_tracks})
of lower-mass models to determine where to divide this track between RGB and AGB).

We interpolate the evolutionary tracks from each stellar model for each group in 
the $\log \Ebind-\log R$ plane, in order to give each model the same number (200) 
of data points, which gives them an equal weight when fitting along the mass axis, and 
with a constant density along their tracks in this plane, which prevents biasing 
the fits towards particular regions in these curves.  The only exception are the 
groups LMR1 and LMR2, which are given 200 data points \emph{together}.
Figure~\ref{fig:be_tracks} shows the model tracks in the LMR, LMA and HM groups 
for a selection of the models in our $Z=0.02$ grid.

\subsection{Selecting the fitting functions}
\label{sec:fitfuncs}

As described above, we choose to find a description for $\log \Ebind$
as a function of $\log M$ and $\log R$.  As a general functional
form, we select the polynomial of an as yet undefined order in both the
instantaneous mass $M$ and radius $R$.  Hence, the general form of our fitting 
function is
\begin{eqnarray}
  && \log \left(\frac{\Ebind}{\mathrm{erg}}\right) \approx
  E_0 +  \nonumber \\
  && \Lambda\left(M_0,M\right) \cdot \sum_{m,r} \alpha_{m,r} \left[\log \left(\frac{M}{M_\odot}\right)\right]^m \left[\log \left(\frac{R}{R_\odot}\right)\right]^r,
  \label{eq:fitfunction}
\end{eqnarray}
with $E_0 = 33.29866$, $\alpha_{m,r}$ the fitting coefficients and $m$ and $r$ integer indices and
exponents.  The factor $\Lambda$ depends on the initial (ZAMS) and instantaneous mass 
and corrects for effects due to wind mass loss in high-mass stars. It is of course equal to
unity for stars that do not experience (significant) mass loss and is discussed 
in more detail in Sect.\,\ref{sec:masslossprocedure}.

Finding a good and accurate prescription for the binding energy then consists
of two parts.  First, we must determine the orders of each of the polynomials 
in Eq.\,\ref{eq:fitfunction}.  In fact, we choose to use a liberal definition
of the word \emph{polynomial}, and rather than selecting the orders, we 
vary the ranges of $m$ and $r$ for which $\alpha_{m,r}$ has non-zero
values in order to find the narrowest ranges that still describe $\Ebind$
with good accuracy, possibly including negative values for $m$ and $r$.
Second, for each of our choices of the ranges of $m$
and $r$, we must find the optimal set of fitting coefficients $\alpha_{m,r}$ 
and quantify the accuracy of the fit, in order to compare to fits that use different 
polynomials.

For the actual fitting of the data points, we use only the \emph{odd-numbered} mass 
models from the grid described in Sect.\,\ref{sec:models} (the masses with odd
values of $i$ in Eq.\,\ref{eq:masses}).
For each selection of the range of $m$ and $r$, we fit the data points, \ie\ 
determine the best set of $\alpha_{m,r}$, using the 
$\chi^2$-minimization-based fitting procedures available in Wolfram Mathematica~7 
\citep{Mathematica7}.  Instead of using Mathematica's 
``goodness-of-fit parameter'', we choose three criteria to compare the accuracies
of the different fits in a quantitative way.

First, we use the newly generated fit function to predict the binding energies
for all mass models and all data points that are used for the fit (\ie\ the 
odd-numbered mass models), and determine
for which fraction of these data points the following condition holds
\begin{equation}
 \left|  \frac{E_\mathrm{bind,fit} - E_\mathrm{bind,mdl}}{E_\mathrm{bind,mdl}} \right| < f,
  \label{eq:accuracy}
\end{equation}
where we choose the value $f=0.1$ for our criterion (note that here we do not use 
the logarithm of $\Ebind$).  The fraction of data points that fulfills 
this criterion for $f=0.1$ is called $\mathbf{\Delta_{10\%}}$.
More specifically, we demand that for at least 
$90\%$ of the data points, the fit has an accuracy of 10\% or better;  in other words, 
$\Delta_{10\%} \geq 90\%$.

Second, we use these fits to compute the envelope binding energies for the \emph{even-numbered} 
mass models in our grid, which were not used to create the fits.  Again, we use the 
condition in Eq.\,\ref{eq:accuracy} for $f=0.1$ to determine the accuracy of the fit, 
and demand that the results are similar to those from the first criterion.

Third, we select the simplest polynomial, \ie\ the narrowest ranges for $m$ 
and $r$ in Eq.\,\ref{eq:fitfunction}, that still pass the first two criteria.

For some groups of data these criteria turn out to be too strict, 
\ie\ there are no reasonable ranges of $m$ and $r$ that
could meet the first criterion while still being sufficiently well behaved 
to also meet the second.  For these groups we are satisfied if the 
fit predicts at least $90\%$ of the data points within $15\%$ of the model value, 
\ie\ we increase $f$ from $0.1$ to $0.15$ for these groups
and demand that $\Delta_{15\%} > 0.9$. For this reason we
report both $\Delta_{10\%}$ and $\Delta_{15\%}$
for all groups in the next section.

\subsection{Compensating for stellar mass loss }
\label{sec:masslossprocedure}

So far, we have considered conservative stellar-evolution models only.  However, we 
find that when stars lose a significant amount of mass due to stellar winds (\ie\ stars 
with masses $\gtrsim 20-30\,\msun$), our fits lose some of their accuracy.
Since wind mass-loss rates are uncertain, and because the assumption of evolution without mass 
loss is unreasonable for the more massive stars in our grids, we introduced the correction
factor $\Lambda$ in Eq.\,\ref{eq:fitfunction}.  The factor is based on a comparison of 
the fit using conservative models, as discussed in the previous section, to grids of 
models which experience mass loss as described in Sect.\,\ref{sec:code}.  
For models in the low-mass grid (LMR/LMA), this correction is
not necessary, and we use $\Lambda=1$.  For the high-mass models (HM), we find that a
correction factor based on the relative amount of mass lost since the zero-age main 
sequence (ZAMS) gives a reasonably good prescription for most models,
restoring the accuracy found for conservative models (see Fig.\,\ref{fig:accur_wind}):
\begin{equation}
  \Lambda_\mathrm{HM}(M_0,M) = 1 + \frac{1}{4} \left( \frac{M_0-M}{M_0} \right )^2,
  \label{eq:masslossfactor}
\end{equation}
where $M_0$ is the ZAMS mass and $M$ is the instantaneous mass of the star.
Note that $\Lambda_\mathrm{HM}$ is independent of metallicity and that it is 
valid across the range of metallicities considered here.

The fitting procedure described in this section was carried out for each of the four groups mentioned 
in Sect.\,\ref{sec:divtracks} and each of the six metallicities listed in 
Sect.\,\ref{sec:models}.  The results of our fits, their accuracies, and their validity for
the case of moderate wind mass loss are discussed in Sect.\,\ref{sec:results}.

\section{Results}
\label{sec:results}

In this section we describe the accuracies of our fits.  In section~\ref{sec:prescriptions}
we present the optimal fit parameters to describe the binding energy as a 
function of basic stellar parameters.  Section~\ref{sec:accuracy} presents
the accuracies of these fits and in Section~\ref{sec:winds} we discuss the
effect of stellar winds on this accuracy.  In Section~\ref{sec:limits}, we
discuss the domain in which our fits are valid.

\subsection{Analytic prescriptions}
\label{sec:prescriptions}

We carried out the fitting of our data points to the fitting function in 
Eq.\,\ref{eq:fitfunction} as described in the Section~\ref{sec:fitting}.
Here, we present the polynomials that best describe the data points,
\ie, the ranges of $m$ and $r$ we used in Eq.\,\ref{eq:fitfunction}, 
and the values for $\alpha_{m,r}$ that best describe the data for those
polynomials.  We do this for each of the four groups of our data set: 
LMR1, LMR2, LMA and HM, as described in Sect.\,\ref{sec:divtracks} and
defined by Tables~\ref{tab:Mlh} and \ref{tab:rgbdivcoeff} and the 
equations in the appendix 
and for each metallicity we consider in this study.

The ranges for $m$, $r$ are variable, since for some groups of models,
the binding energy can be described in fewer terms than for others.  The extreme
values used are 0 and 20 for $m$ and $-5$ and 20 for $r$.
We provide the table of coefficients in electronic form, listing $m$, $r$ and $\alpha_{m,r}$
for all cases where $\alpha_{m,r}$ is non-zero, as well as the contents of Tables~\ref{tab:Mlh} 
and \ref{tab:rgbdivcoeff}.  In addition, we provide \texttt{Fortran} routines which can read these 
data files and compute the envelope binding energy as a function of metallicity, mass, radius, 
and evolutionary phase of the star (RGB or AGB) in \citet{edata}. 

\begin{figure}  
  \includegraphics[angle=-90,width=\columnwidth]{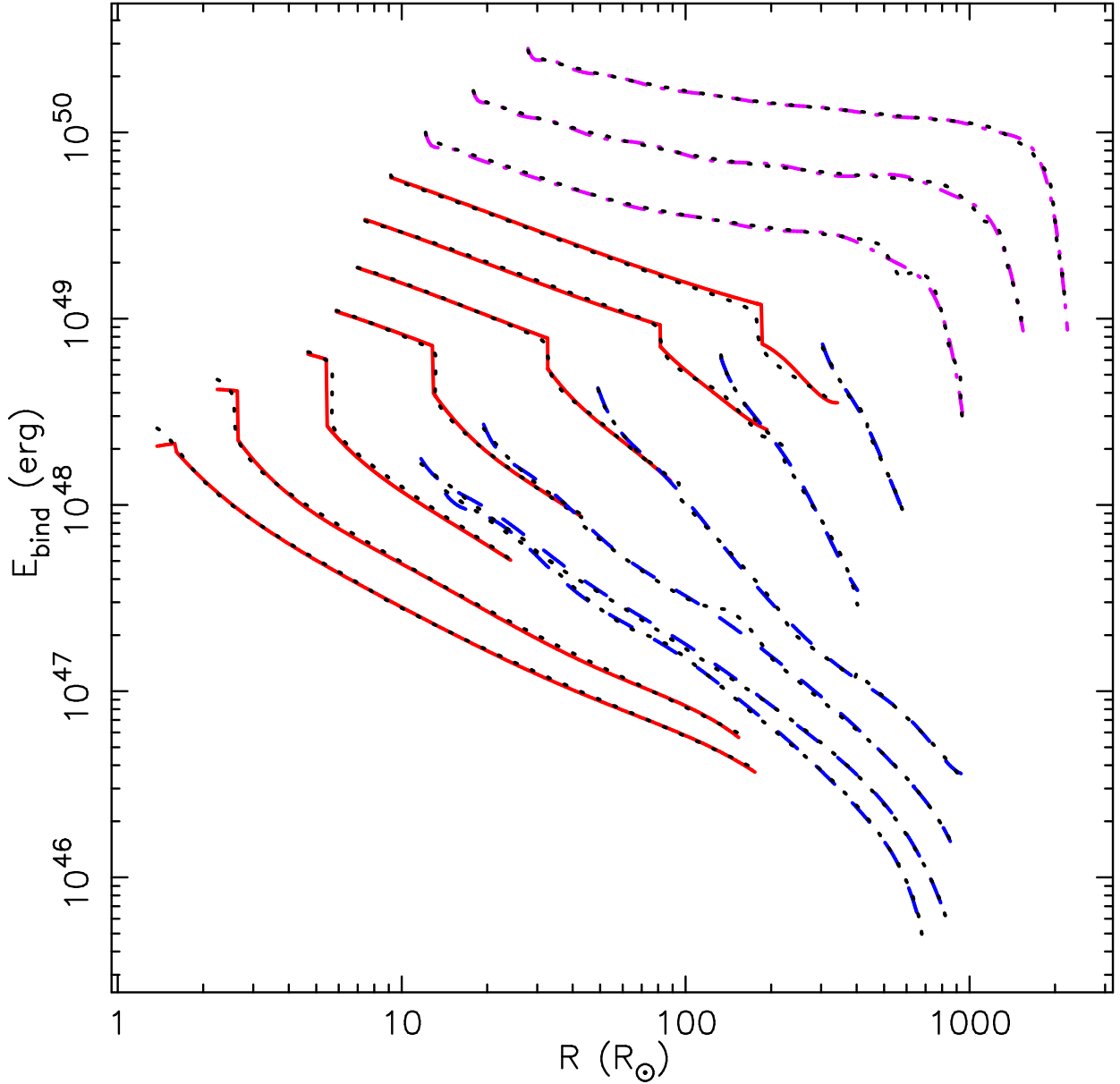}
  \figcaption{
    Envelope binding energy as a function of stellar radius, for a selection of models
    with masses of 0.91, 1.36, 2.04 3.05, 4.57, 6.84, 10.22, 15.3, 22.9 and 34.2\,$\msun$, 
    and $Z=0.02$.  The line styles and colours indicate the same evolutionary phases as in 
    Fig.\,\ref{fig:hrd_tracks} and are used for the results obtained by using our fit.
    The black dotted lines show the original stellar-evolution models, which overlap with the fits
    in most places.  RGB and AGB phases are
    disconnected here, and the lowest-mass and the three highest-mass models do not have an 
    AGB phase.
    \label{fig:be_tracks}
  }
\end{figure}

\begin{figure}  
  \includegraphics[angle=-90,width=\columnwidth]{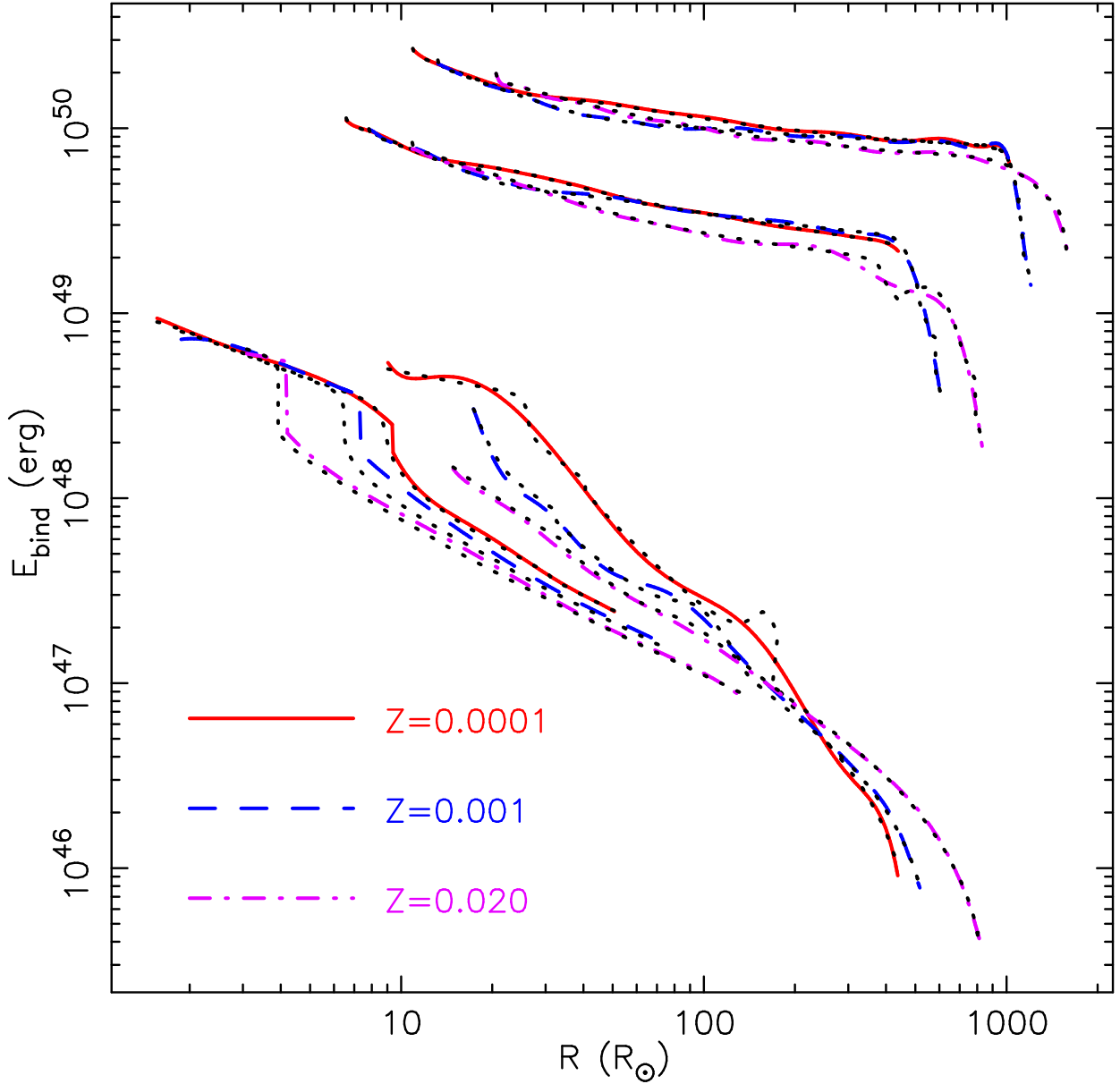}
  \figcaption{
    Envelope binding energy as a function of stellar radius, for models with masses of
    $M=1.79$, $13.4$ and $26.2\,\msun$ and for metallicities of $Z=10^{-4}$, $10^{-3}$ 
    and $0.02$.  The line styles and colours indicate the value of Z as indicated in 
    the figure.  Coloured lines represent fitting results, while black dotted lines 
    show the detailed models.  The four groups of lines are from lower left to upper 
    right: $1.79\,\msun$ RGB, $1.79\,\msun$ AGB, $13.4\,\msun$ and $26.2\,\msun$.  Within 
    each group, stars with a lower metallicity have a higher binding energy, except
    near the end of the AGB or high-mass tracks.
    \label{fig:be_tracks_Zs}
  }
\end{figure}

Figure~\ref{fig:be_tracks} shows a comparison between the detailed stellar-evolution 
models and fits for selected models with a range of masses from our $Z=0.02$ grid.
Figure~\ref{fig:be_tracks_Zs} shows the dependence of the envelope binding energy on
metallicity, by comparing models with different metallicities for three different 
masses.

\subsection{Accuracy of the results}
\label{sec:accuracy}

Table~\ref{tab:accuracies} lists the accuracies of our fits for the 
envelope binding energy for each of the
metallicities considered, and for each of the model groups defined in 
Sect.\,\ref{sec:divtracks}.  To express the accuracy of each fit,
we list the percentages of the fitted data points that fall within $10\%$ and 
$15\%$ of the model values ($\Delta_{10\%}$ and $\Delta_{15\%}$).

\begin{table}
  \caption{
    Accuracies of our fits, for each metallicity considered and
    each group defined in Sect.\,\ref{sec:divtracks}.
    For an explanation of the $\Delta$'s, see Sect.\,\ref{sec:fitfuncs}
    and Eq.\,\ref{eq:accuracy}.
    \label{tab:accuracies}
  }
  \centering
  \begin{tabular}{cccccc}
    $Z$  & group & \multicolumn{2}{c}{$\Delta_{10\%}$} & \multicolumn{2}{c}{$\Delta_{15\%}$} \\
         &       & odd & even & odd & even \\ 
    \hline
    $10^{-4}$ & LMR  & 97.6\% & 97.7\% & 98.9\% & 99.1\%  \\
    $10^{-4}$ & LMA  & 86.6\% & 86.4\% & 94.8\% & 94.8\%  \\
    $10^{-4}$ & HM   & 94.8\% & 93.6\% & 96.4\% & 94.6\%  \\ 
    \multicolumn{6}{l}{} \\
    $10^{-3}$ & LMR  & 92.0\% & 90.4\% & 95.0\% & 93.6\%  \\
    $10^{-3}$ & LMA  & 85.3\% & 84.3\% & 95.2\% & 92.5\%  \\
    $10^{-3}$ & HM   & 91.1\% & 90.0\% & 95.5\% & 93.4\%  \\ 
    \multicolumn{6}{l}{} \\
    $0.01$    & LMR  & 95.7\% & 93.7\% & 97.1\% & 96.5\%  \\
    $0.01$    & LMA  & 86.3\% & 87.2\% & 95.9\% & 95.0\%  \\
    $0.01$    & HM   & 92.4\% & 90.0\% & 96.2\% & 96.1\%  \\
    \multicolumn{6}{l}{} \\
    $0.015$   & LMR  & 95.0\% & 95.9\% & 96.6\% & 97.0\%  \\
    $0.015$   & LMA  & 91.4\% & 90.6\% & 97.8\% & 96.4\%  \\
    $0.015$   & HM   & 90.0\% & 90.3\% & 95.5\% & 95.8\%  \\ 
    \multicolumn{6}{l}{} \\
    $0.02$    & LMR  & 94.3\% & 92.7\% & 96.4\% & 96.4\%  \\
    $0.02$    & LMA  & 97.1\% & 91.9\% & 99.3\% & 96.9\%  \\
    $0.02$    & HM   & 92.0\% & 91.7\% & 96.6\% & 96.1\%  \\ 
    \multicolumn{6}{l}{} \\
    $0.03$    & LMR  & 95.0\% & 94.7\% & 97.1\% & 96.4\%  \\
    $0.03$    & LMA  & 97.0\% & 90.5\% & 98.7\% & 92.5\%  \\
    $0.03$    & HM   & 91.1\% & 90.4\% & 96.3\% & 95.8\%  \\ 
  \end{tabular}
\end{table}

\begin{figure*} 
  \includegraphics[angle=-90,width=\textwidth]{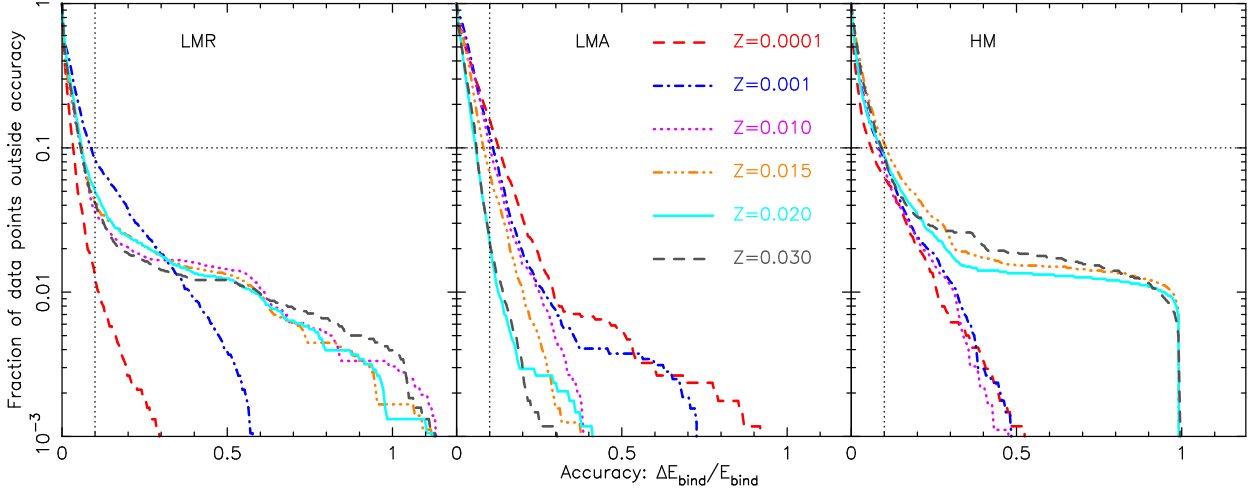}
  \figcaption{
    Cumulative histograms for the accuracies (as defined in Eq.\,\ref{eq:accuracy})
    in our three different groups of models for all six metallicities in our grid.
    On the vertical axis, we plot the fraction of model data points that fall 
    \emph{outside} the accuracy quoted on the horizontal axis, hence
    a fraction of 0.1 means that 90\% of our data points have that accuracy.
    \emph{Left panel:} low-mass RGB, \emph{middle panel:} low-mass AGB, \emph{right panel:} high mass.
    The different line styles and colours indicate different metallicities, as indicated in the middle panel, 
    and the vertical and horizontal dotted black lines indicate the points where the accuracy is 10\% and where 90\%
    of our model data points are included (10\% excluded), respectively, and their intersection marks 
    the point $\Delta_{10\%} = 90\%$.
    \label{fig:accur_nowind}
  }
\end{figure*}

We separately list the accuracies for the odd-numbered stellar-evolution models, 
which were used to produce the fits, and the even-numbered models, which have
masses that lie between those of the odd-numbered models and are used only for 
verification.  Note that the percentages refer to the actual (absolute) value
of the binding energy, not its logarithm.
The same results are presented in Fig.\,\ref{fig:accur_nowind}, as cumulative histograms
of the fraction of data points that lie \emph{outside} a given accuracy.

As these data demonstrate, all of our fits meet the criteria
outlined in the previous section, for $f=0.15$ in Eq.\,\ref{eq:accuracy}. In fact, all but 
three of the fits also meet the stricter criteria for $f=0.1$. 
For $Z \leq 0.01$, the features in the LMA models vary irregularly from one mass 
model to the next, so that only for our low-metallicity AGB models, the $f=0.1$ 
criteria are not satisfied.  \citet{2010ApJ...720.1752P} show that, for the cases
where a common envelope (CE) leads to a merger, only 17\% of the CEs are initiated on
the AGB, against 83\% on the RGB, so that the lower accuracy for our LMA results 
probably has a somewhat smaller impact than suggested by the numbers in the table
alone.

\subsection{The effect of stellar mass loss}
\label{sec:winds}

\begin{figure*} 
  \centering
  \includegraphics[angle=-90,width=0.67\textwidth]{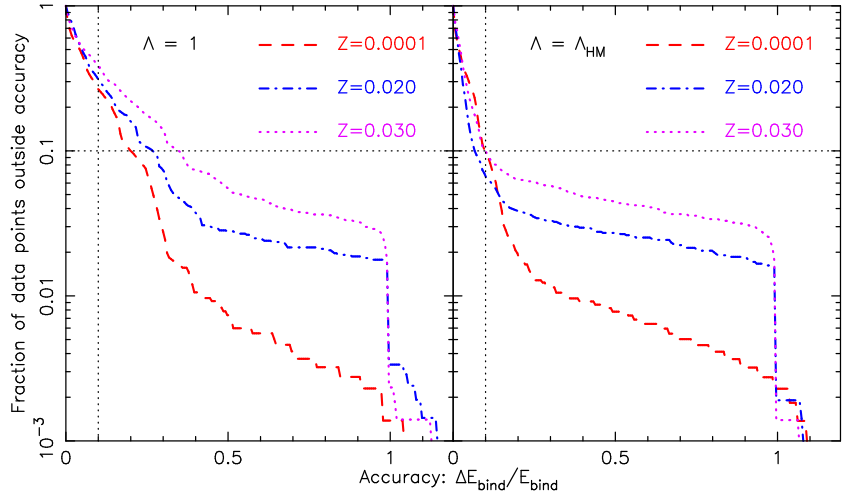}
  \figcaption{
    Cumulative histograms for the accuracies (as defined in Eq.\,\ref{eq:accuracy})
    in our high-mass models HM with wind mass loss, for a selection of metallicities in our grid.
    \emph{Left panel:} no wind-correction factor applied ($\Lambda=1$), 
    \emph{right panel:} with $\Lambda=\Lambda_\mathrm{HM}$, as defined in Eq.\,\ref{eq:masslossfactor}, applied.
    The different line styles and colours indicate different metallicities, as indicated in the panels, 
    and the vertical and horizontal dotted black lines indicate the points where the accuracy is 10\% and where 90\%
    of our model data points are included (10\% excluded), respectively.
    \label{fig:accur_wind}
  }
\end{figure*}

The accuracies in the previous section apply to the prediction of the binding energies of 
stars that have not suffered mass loss due to stellar winds. 
Here, we compare the outcome of our fits, based on conservative models, to grids of stellar
models that undergo wind mass loss as described in Sect.\,\ref{sec:code}.  Table~\ref{tab:windaccuracies}
lists the accuracies for this comparison for the two extreme metallicities in our grid 
($Z=10^{-4}$ and $0.03$) as well as ``solar'' metallicity ($Z=0.02$),
expressed as $\Delta_{10\%}$ and $\Delta_{15\%}$ (see Sect.\,\ref{sec:fitfuncs}). We see that when 
the factor $\Lambda$ in Eq.\,\ref{eq:fitfunction} is ignored, the LMR grid is hardly 
affected compared to Table~\ref{tab:accuracies}, whereas the accuracies in the LMA and 
especially the HM grids suffer appreciably.  As described in Sect.\,\ref{sec:masslossprocedure}, 
the correction factor $\Lambda$ defined in Eq.\,\ref{eq:masslossfactor} improves the 
accuracies for the high-mass models significantly. This improvement is also shown
in Fig.\,\ref{fig:accur_wind}.
For the LMA group, we found no simple correction factor that is independent of metallicity.
Since the drop in accuracy due to winds is not as large as for the LM models, we decided to leave 
these fits uncorrected.

\begin{table}
  \caption{
    Accuracies of our fits 
    as in Table~\ref{tab:accuracies},
    but 
    applied to models with stellar wind,
    with $\Lambda=1$ and $\Lambda=\Lambda_\mathrm{HM}$ (Eq.\,\ref{eq:masslossfactor}). 
    \label{tab:windaccuracies}
  }
  \centering
  \begin{tabular}{cccccccc}
          &      && \multicolumn{2}{c}{$\Lambda=1$,} && \multicolumn{2}{c}{$\Lambda=\Lambda_\mathrm{HM}$,} \\
    Group & $Z$  && $\Delta_{10\%}$ & $\Delta_{15\%}$                  && $\Delta_{10\%}$ & $\Delta_{15\%}$ \\
    \hline
    LMR & $10^{-4}$ && 98.8\% & 99.3\%    && & \\
        & $0.02$    && 94.3\% & 96.4\%    && & \\
        & $0.03$    && 93.7\% & 96.4\%    && & \\ 
    \multicolumn{6}{l}{}  \\    
    LMA & $10^{-4}$ && 77.3\% & 87.4\%    && & \\
        & $0.02$    && 75.5\% & 87.5\%    && & \\
        & $0.03$    && 68.3\% & 76.6\%    && & \\
    \multicolumn{6}{l}{} \\
    HM  & $10^{-4}$ && 69.7\% & 86.4\%    && 92.1\% & 96.1\%  \\
        & $0.02$    && 54.7\% & 63.1\%    && 87.2\% & 89.5\%  \\
        & $0.03$    && 60.1\% & 70.7\%    && 87.0\% & 89.8\%  \\  
  \end{tabular}
\end{table}

The advantage of the $\Lambda$ factor is that our prescription for the binding energy can be used with arbitrary mass-loss 
prescriptions and still provide the accuracies listed in Table~\ref{tab:windaccuracies}.  However, we
find that for very strong mass loss, even this correction factor can no longer describe our models
accurately.  Our results show that when a star loses more than $\sim20\%$ of its ZAMS mass, our fits
become increasingly inaccurate.  For each metallicity, this implies an upper limit in mass
for which these fits can be used.  We did not include the results of 
more massive models in 
Table~\ref{tab:windaccuracies} and discuss these limitations
in more detail in Sect.\,\ref{sec:limits}.

\subsection{Limits on the domain of the fits}
\label{sec:limits}

Here we note the limitations on the applicability of the fit functions.
Outside these limits we do not provide fits and hence cannot estimate accuracies.

In the absence of mass loss, Table~\ref{tab:accuracies} displays the accuracies of our fits for any
star with a mass $0.8\,\msun \leq M \leq 100\,\msun$, the full range of masses we computed.
However, when mass loss is present, we note that the parameter $\Lambda_\mathrm{HM}$ is only 
effective when stars have lost less than $\sim 20\%$ of their ZAMS masses.
For the wind prescriptions used in this study, this limit can be translated to an upper mass limit
of approximately $50\,\msun$ for $Z=0.02$ and $Z=0.03$, and about $75 \msun$ for $Z=10^{-4}$.

In terms of radius, we have considered each model from the end of the MS 
to the star's maximum radius on the AGB, excepting only phases of radial shrinkage
--- most notably on the horizontal branch, so that any radius that may correspond to the 
initiation of a CE in a binary is covered in our study.

The values of the metallicity considered here are discrete, and we do not provide accuracies 
for $Z<10^{-4}$ and $Z>0.03$.  Analogous to metallicities that lie between those used in
this study, for metallicities outside this range the reader can either choose to use the value 
for $Z$ that lies closest to the desired value, or to use an extrapolation scheme.

\section{Discussion}
\label{sec:discussion}

In this section we discuss our results qualitatively, compare our fits to previous attempts to
model the binding energy, note the possible effects of the choice of core definition, mixing-length ratio, and
strong winds on the accuracy and applicability of our results.

\subsection{Qualitative examination of the results}
\label{sec:qualdisc}

Figures~\ref{fig:be_tracks} and \ref{fig:be_tracks_Zs} show examples of the envelope binding energies of a selection
of detailed stellar-evolution models (black dotted lines) compared with the results of our fits (coloured lines).
These figures indicate that the fitting function we defined in Eq.\,\ref{eq:fitfunction} describes the 
binding energy fairly well.  A full picture of the accuracy of our fits, when compared to our models, can be read
from Table~\ref{tab:accuracies} and Fig.\,\ref{fig:accur_nowind}, and from Table~\ref{tab:windaccuracies} and 
Fig.\,\ref{fig:accur_wind}, for models with conservative evolution and models that undergo wind mass loss, respectively.

We want to emphasise here the limitation of our results for stars that undergo substantial wind mass loss.
When stars lose more than $\sim 20\%$ of their ZAMS mass, the binding energy is no longer easily related to that of a 
star of the same mass but without a mass-loss history. This is the result of an interplay between the different effects 
of wind mass loss on both the mass and structure of the envelope. Strong mass loss ultimately reduces the
envelope \emph{mass}, which decreases the binding energy compared to a conservative star of the same total mass and radius. 
However, our models indicate that the envelope \emph{structure} is also affected by mass loss, in a way that counteracts the
first effect.  We find that this results in a larger binding energy than that of a conservative star, unless the mass loss becomes very significant.  
Hence, our compensation factor $\Lambda_\mathrm{HM}$ in Eq.\,\ref{eq:masslossfactor}, which is greater than unity, 
gives reasonably good results for small amounts of mass loss, but becomes gradually less accurate when more than 
$\sim 20\%$ of the ZAMS mass of the star is lost.  We note that for our choice of wind mass-loss prescription, this 
happens for stars with masses larger than $\gtrsim 50\,\msun$ for $Z=0.02, 0.03$ and for $M \gtrsim 75\,\msun$ for 
$Z=10^{-4}$, hence the most massive stars in our grids.  These upper limits will be decreased when stronger winds are used
in the stellar-evolution models.  However, as we shall discuss in Sect.\,\ref{sec:core_envelope}, the dominant source 
of uncertainty for these high-mass stars is the definition of the core-envelope boundary.

\subsection{Dependence on the mixing-length ratio}
\label{sec:mlr}

The choice of the mixing length relative to the pressure scale height will directly influence the radius of a model star with 
a deep convective envelope.  \citet{2000A&A...360.1043D} looked at the effect of this parameter and found a weak dependence
for the binding energy, with a higher mixing-length ratio translating into a slightly lower binding energy on the AGB.
We compared our fits based on our standard assumption $l/H_\mathrm{p} = 2.0$ to a few test models which were computed with 
$l/H_\mathrm{p} = 1.5$ and found a similar effect: 
a drop in accuracy almost exclusive to the AGB, resulting in $\sim$15--20\% more data points outside the 10\% accuracy range.

\subsection{Definition of the core-envelope boundary}
\label{sec:core_envelope}

The binding energy of a stellar envelope is computed by integrating the different energy sources 
(thermal, gravitational, etc.) from the core-envelope
boundary to the surface of the star (Eq.\,\ref{eq:Ebind}).  Hence, the value of the binding energy will depend on the 
definition of the core mass $M_\mathrm{c}$, which we chose to define as the mass coordinate at which the hydrogen 
abundance reaches 10\%.  \citet{2001A&A...369..170T} consider several reasonable choices for this definition, one of
which is identical to the definition used in our study (the central region of the star with $X<0.1$).  
If we follow these authors and discard the extreme cases (first 
and last row) in their Table~1 (their $\lambda_\mathrm{b}$ is almost identical to our $\lam$ defined in 
Eq.\,\ref{eq:lambda}), we see that while for low-mass stars different definitions result in similar values for 
$M_\mathrm{c}$ and hence the binding energy, for more massive stars the results can vary quite appreciably. For a
$20\,\msun$ star, the variation in $E_\mathrm{bind}$ can even be more than an order of magnitude.  They also note that our choice of the 
definition of the core gives the lowest value for $\lam$, hence the highest value for the binding
energy, among the three options considered.

We therefore conclude that the definition of the core mass will have little or no influence on our 
results for low-mass stars, with masses $\lesssim 7 - 10\,\msun$.  Since these stars provide the majority of 
common-envelope donors in binaries \citep{MvdS_Mykonos}, most \emph{instances} of CEs in nature will be little affected,
as will the results for low-mass compact binaries, such as double white dwarfs, cataclysmic variables and low-mass
X-ray binaries.  On the other hand, for high-mass stars, the mass of the stellar core will depend strongly on how
the core-envelope boundary is defined.  
  In fact, in reality the exact remnant mass after the CE will depend on the response of both the donor star 
  and the orbit to the spiral-in, and hence also on the properties of the binary companion 
  \citep{2010ApJ...719L..28D,Ivanova10}.  Therefore, for these massive stars no unique core mass can be defined 
  and detailed stellar-structure and binary-evolution models are needed to determine a self-consistent outcome 
  of a CE. The uncertainty thus introduced is much larger than the uncertainty we find
  in our fitting results for high-mass stars with strong winds described in Sect.\,\ref{sec:qualdisc}.

Finally, we would like to remark that our choice of the core-envelope-boundary definition is identical to that used
for the fits that resulted in the stellar-evolution prescriptions in the \texttt{SSE} and \texttt{BSE} codes
\citep{2000MNRAS.315..543H,2002MNRAS.329..897H}, which form the basis of a number of population-synthesis codes,
such as \texttt{StarTrack} \citep{2008ApJS..174..223B} and the latest version of \texttt{SeBa} 
\citep{2001A&A...365..491N,Silvia_SeBa}.  Hence, even if there is no unique definition of the core mass, our 
fits will provide consistent prescriptions when implemented in the most-commonly used population-synthesis codes.

\subsection{The envelope-structure parameter $\lam$}
\label{sec:lambda}

\begin{figure}  
  \includegraphics[angle=-90,width=\columnwidth]{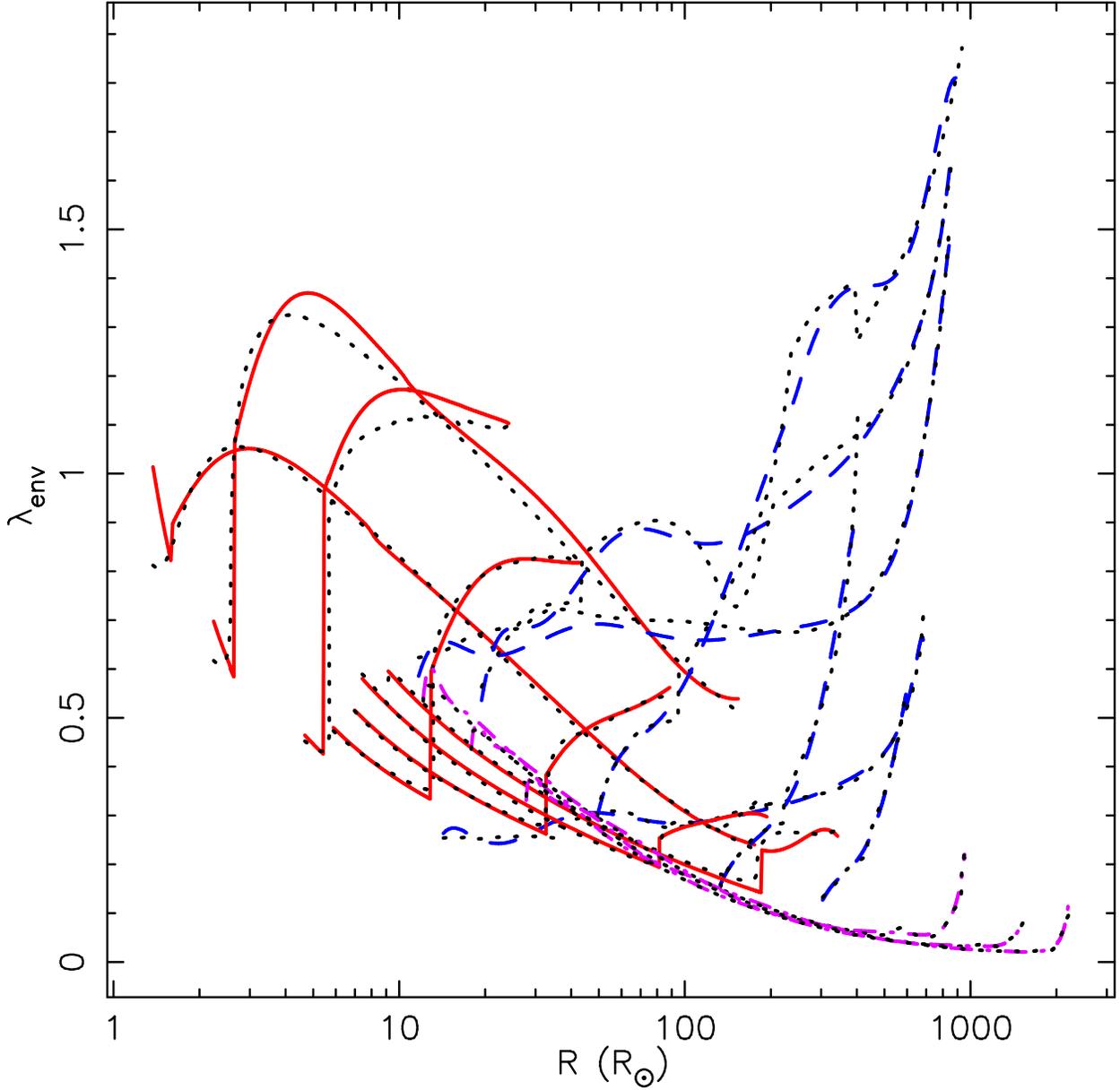}
  \figcaption{
    Typical values of $\lam$ for a selection of models in our $Z=0.02$ grid.  Shown are the
    LMR (solid red lines) and LMA (dashed blue lines) for low-mass models and HM (dash-dotted magenta lines)
    for the high-mass model.  Masses shown are $1.04\,\msun$ (long, high LMR; short, low LMA), $4.57\,\msun$
    (short, lower LMR; long, high LMA) and $15.3\,\msun$ (HM track).  In all cases, coloured lines indicate
    results from our fits, dotted, black lines show the detailed stellar model.
    \label{fig:lambdas}
  }
\end{figure}

The purpose of this study is to provide fits of the binding energy of stellar envelopes, based on basic stellar
parameters, so that they can be used to treat CEs in population-synthesis simulations, where detailed
stellar-structure models are not available.
So far, such codes have often used the so-called \emph{envelope-structure parameter}, $\lam$, defined by
\begin{equation}
  \Ebind = \frac{G M M_\mathrm{env}}{R \lam},
  \label{eq:lambda}
\end{equation}
to compute the binding energy \citep{1984ApJ...277..355W,1987A&A...183...47D}, instead of Eq.\,\ref{eq:Ebind}.  
In such a case, one needs to assume a value for $\lam$.  For example,
\citet{2000A&A...360.1011N} and \citet{2002MNRAS.329..897H} use $\lam = 0.5$, while 
\citet{2008ApJS..174..223B} choose $\alpha_\mathrm{CE} \lam = 0.5, 1.0$.  Since $\lam$ has been used
extensively, it is useful to give an indication here of our results expressed in terms of this 
parameter, and to compare them to the choices above.  In Figure~\ref{fig:lambdas}, we show the values of $\lam$ for 
three different stellar-evolution models and the results of our fits, converted to $\lam$.  This figure alone 
indicates that $\lam$ is far from constant and will vary for stars of different masses and different 
evolutionary phases.  \citet{2000A&A...360.1043D} and \citet{2006A&A...460..209V} give more extensive 
examples of the variation of this parameter.  In addition, 
\citet{MvdS_Mykonos} show that, for $Z=0.02$, the assumption of $\alpha_\mathrm{CE} \lam = 1.0$ implicitly 
assumes that $\alpha_\mathrm{CE} > 1$ for roughly 60\% of the common envelopes in a typical stellar population, 
while for some cases $\alpha_\mathrm{CE} > 10$ is implied.  These examples strongly indicate that a better 
prescription for the envelope binding energy is needed.

The reason why we chose to develop an expression for the envelope binding energy, and not $\lam$, may now become
clear.  The parameter was introduced in order to estimate the binding energy when it could not
be computed otherwise.  From now on, a reasonably accurate prescription for $\Ebind$ is available and there
is no longer any need to use the parameter $\lam$.

\subsection{Comparison with previous work}
\label{sec:comparison}

Previous studies that provided a simple prescription for the binding energy have been limited to tabulations of 
$\lam$ \citep{2000A&A...360.1043D} for stars of different masses and different evolutionary stages.
However, while our paper was in preparation, a paper was published by \citet{2010ApJ...716..114X} largely 
with the same idea as our study: to create simple and fast prescriptions that can provide the envelope 
binding energy in order to treat CEs in population-synthesis codes. 

The main points where our study improves
on the latter paper concern the range of applicability.  Firstly, whereas 
\citet{2010ApJ...716..114X} provide fits for a selection of 14 discreet masses between $1\,\msun$ and $20\,\msun$, 
we use the stellar mass as a continuous parameter so that it can take any value between $0.8\,\msun$ and $100\,\msun$.
Hence, the user does not have to interpolate between masses, which may not be a trivial process. For example, while stars of $1.0\,\msun$ 
and $1.5\,\msun$ reach a radius of $100\,\rsun$ on the RGB, a $2.0\,\msun$ star does not.  It may therefore be impossible
to obtain the binding energy for a $1.5\,\msun$ star at that radius on the RGB by interpolating the results of
the $1.0\,\msun$ and $2.0\,\msun$ models in a discrete grid.
Secondly, they provide two choices of metallicity ($Z=0.001$ and $Z=0.02$), whereas we include six different,
albeit still discreet, values, ranging from $Z=10^{-4}$ to $Z=0.03$.  
We believe that our grid of metallicities is sufficiently dense that
the nearest value can be used without too much loss in accuracy.
Thirdly, whereas their study includes a fixed choice of stellar winds, our fits allow for the user's choice 
of wind prescription (although the accuracy we present is valid only within certain limits as discussed in 
Sect.\ref{sec:qualdisc}).

In addition to wider a applicability, our paper presents a more extensive error analysis.
As we have shown in Sect.\,\ref{sec:qualdisc}, our accuracies are dominated by the uncertainty in wind 
mass loss for massive stars, by the uncertainty in the separation between core and envelope at the end of a 
CE, or even by the intrinsic problems in determining this separation a priori for stars with masses of 
$\sim 10-20\,\msun$ and up (Sect.\,\ref{sec:core_envelope}).  These systematic errors are discussed 
by \citet{2010ApJ...716..114X} as well.
However, for the majority of common envelopes the donor
mass is $\lesssim 10\,\msun$, so that these external uncertainties are less important. For these cases the
uncertainty in the fits will dominate, which we present explicitly in this paper.

On the other hand, the study by \citet{2010ApJ...716..114X} has two advantages over ours. First, their work covers
the HB for all masses they consider. While we include all stages of stellar evolution where a CE could be initiated, knowledge 
of the envelope binding energy on the HB for low-mass stars, which cannot initiate a CE, 
may be useful for \eg\ collisions. Second, their prescriptions are simpler than ours, making them easier to implement.

\section{Summary and conclusions}
\label{sec:conclusions}

We provide analytical fits capable of predicting the envelope binding energy of stars with masses between $0.8\,\msun$
and $100\,\msun$ on the giant branches, for any given mass and radius, and for six discrete choices of the metallicity.
These fits are based on detailed stellar-evolution models.  In addition, we define an ad-hoc correction factor that takes
into account wind mass loss.  We provide electronic data files with the fitting coefficients and \texttt{Fortran} routines 
and that are ready to use these files \citep{edata}.

We find that the accuracy of our fits is better than 15\% for 90\% of our model data 
points in all cases and better than 10\% for 90\% of the data points for most cases, when fractional mass loss since the ZAMS 
is less than $\sim 20\%$.  For low-mass stars, the true accuracy will be close to the value we quote, whereas for high-mass
stars the uncertainty in the determination of the core-envelope boundary will probably be the limiting factor.
We conclude that our fits allow population-synthesis codes and other environments where detailed stellar models are not 
available to compute envelope binding energies for \eg\ common envelopes quickly and accurately.
In addition, the results presented here are more widely applicable and more accurate than previous prescriptions.

\acknowledgements{
  We would like to thank P.P.\ Eggleton for making his binary-evolution code \texttt{ev} available to us.  
  A.J.L.\ acknowledges support from the Northwestern Undergraduate Research Grant and the Weinberg 
  College Summer Grant programs.
  MvdS acknowledges support from a CITA National Fellowship to the University of Alberta.
  The group work was further supported by a NSF CAREER Award AST-0449558 to VK. Simulations were performed 
  using the high-performance computing cluster available to the Theoretical Astrophysics group at Northwestern 
  funded through a past NSF MRI grant to VK.
}

\bibliography{ubind}{}
\bibliographystyle{apj}

\end{document}